# Formation Of A Cold Antihydrogen Beam in AEGIS For Gravity Measurements

G. Testera[a], A.S. Belov[b], G. Bonomi[c], I. Boscolo[d], N. Brambilla[d], R. S. Brusa[e], V.M. Byakov[f], L. Cabaret[g], C. Canali[h], C. Carraro[a], F. Castelli[d], S. Cialdi[d], M. de Combarieu[i], D. Comparat[g], G. Consolati[j], N. Djourelov[l], M. Doser[m], G. Drobychev[w], A. Dupasquier[j], D. Fabris[v], R. Ferragut[j], G. Ferrari[o], A. Fischer[h], A. Fontana[c], P. Forget[g], L. Formaro[d], M. Lunardon[v], A. Gervasini[d], M.G. Giammarchi[d], S.N. Gninenko[b], G. Gribakin[p], R. Heyne[h], S.D. Hogan[q], A. Kellerbauer[h], D. Krasnicky[u], V. Lagomarsino[a], G. Manuzio[a], S. Mariazzi[e], V.A. Matveev[b], F. Merkt[q], S. Moretto[v], C. Morhard[h], G. Nebbia[v], P. Nedelec[n], M.K. Oberthaler[r], P. Pari[i], V. Petracek[u], M. Prevedelli[o], I. Y. Al-Qaradawi[z], F. Quasso[j], O. Rohne[t], S. Pesente[v], A. Rotondi[c], S. Stapnes[t], D. Sillou[n], S.V. Stepanov[g], H. H. Stroke[s], G. Tino[o], A. Vairo[d], G. Viesti[v], H. Walters[p], U. Warring[h], S. Zavatarelli[a], A. Zenoni[c], D.S. Zvezhinskij[g] (AEGIS Proto-Collaboration)

[a] *Istituto Nazionale di Fisica Nucleare and Department of Physics, University of Genoa, via Dodecaneso 33, 16146 Genova, Italy*
[b] *Institute for Nuclear Research of the Russian Academy of Sciences, Moscow 117312, Russia*
[c] *Istituto Nazionale di Fisica Nucleare and Department of Physics, University of Pavia, via Bassi 6, 27100 Pavia, Italy*
[d] *Istituto Nazionale di Fisica Nucleare and Department of Physics, University of Milan, via Celoria 16, 20133 Milano, Italy*
[e] *Department of Physics, University of Trento, via Sommarive 14, 38100 Povo, Italy*
[f] *Institute for Theoretical and Experimental Physics, Moscow 117218, Russia*
[g] *Laboratoire Aimé Cotton, CNRS, Univ Paris-Sud, Bât. 505, 91405 Orsay, France*
[h] *Max Planck Institute for Nuclear Physics, Postfach 103980, 69029 Heidelberg, Germany*
[i] *Laboratoire de Basse Tempe´rature du SPEC/DRECAM/DSM, CEA Saclay, 91191 Gif-sur-Yvette Cedex-France*
[l] *Institute for Nuclear Research and Nuclear Energy, 72 Tzarigradsko Chaussee, 1784 Sofia, Bulgaria*
[m] *Department of Physics, CERN, 1211 Gene`ve 23, Switzerland*
[n] *Lab. d'Annecy-le-Vieux de Phys. des Particules, 9 Chemin de Bellevue, B.p. 110, 74941 Annecy Cedex, France*
[j] *Department of Physics, Politecnico di Milano, Piazza L. da Vinci 32, 20133 Milano, Italy.*
[o] *Istituto Nazionale di Fisica Nucleare and Department of Physics, University of Florence and LENS, CNR-INFM, via Sansone 1, 50019 Firenze, Italy*
[p] *Department of Applied Mathematics and Theoretical Physics, Queen's University, University Road, Belfast BT7 1NN, United Kingdom*
[q] *Laboratorium für Physikalische Chemie, ETH Zürich, Zürich 8093, Switzerland*
[r] *Kirchhoff Institute of Physics, University of Heidelberg, Im Neuenheimer Feld 227, 69120 Heidelberg, Germany*
[s] *Department of Physics, New York University, 4 Washington Place, New York, NY 10003, USA*

[t] University of Oslo, Department of Physics, Oslo, Norway
[u] Czech Technical University in Prague, Faculty of Nuclear Sciences and Physical Engineering
[v] Istituto Nazionale di Fisica Nucleare and Department of Physics, University of Padova,
Via Marzolo 8, 35131 Padova - Italy
[z] Qatar University, Doha, Qatar P. O. Box 2713, Doha, Qatar
[w] Institute for Nuclear Problems of the Belarus State University, 11, Bobruiskaya Str., Minsk 220030, Belarus

**Abstract.** The formation of the antihydrogen beam in the AEGIS experiment through the use of inhomogeneous electric fields is discussed and simulation results including the geometry of the apparatus and realistic hypothesis about the antihydrogen initial conditions are shown. The resulting velocity distribution matches the requirements of the gravity experiment. In particular it is shown that the inhomogeneous electric fields provide radial cooling of the beam during the acceleration.

**Keywords:** Antihydrogen, Equivalence Principle, Rydberg atoms.
**PACS:** 04.80.Cc, 32.80.Ee, 34.70.+e, 36.10.-k, 37.10.Ty

# INTRODUCTION

The equivalence principle is a foundation of General Relativity and a large experimental effort is placed in testing its consequences in all possible fields: this research activity includes tests about the equality of the inertial and gravitational mass, the universality of the free fall, the search for non Newtonian corrections to the gravitational law, the measurement of the gravitational red shift, the search for time variation of the fundamental constants. Measurements about the equality of the inertial and gravitational mass of different bodies began well before the general relativity (they started with Galileo and Newton) and then continued over the years with larger and larger precisions: sensitivity of the order of $10^{-15}$ and $10^{-18}$ are expected from modern space experiments [1] while laboratory torsion balance setups have already reached an accuracy of 1 part in $10^{13}$ [2].

While all these tests concern macroscopic bodies, a sensitivity to the gravitational acceleration g of the order of $10^{-10}$ has been obtained with cold atoms in atomic fountains [3].

The important point is that all these measurements (with macroscopic bodies or cold atoms) are performed on matter system: there are no direct measurements about the validity of the principle of equivalence for antimatter. The validity of the equivalence principle for antimatter is extrapolated from the matter results or it is inferred using indirect arguments (several of them are controversial and model dependent [4]). Particularly interesting is that some quantum gravity models leave room for possible violations of the equivalence principle for antimatter [4].

The AEGIS experiment has been proposed [5,6] at the CERN Antiproton Decelerator to directly measure the Earths gravitational acceleration g on a beam of cold antihydrogen. The basic setup of the experiment should allow one to reach an accuracy of 1% and the experiment is designed to allow higher precision measurements through radial cooling of the beam.

The gravitational acceleration g will be obtained by detecting the vertical deflection of the antihydrogen beam flying horizontally with a velocity of a few 100 m/s for a path length of about 1 meter. The small deflection (few tens of μm) will be measured using two material gratings coupled to a position sensitive detector working as Moiré deflectometer in the classical regime. A device of this type with an additional grating in place of the position sensitive detector has been already operated with atoms [7] by members of the collaboration.

While details about the antihydrogen formation in AEGIS and the Moiré device are reported elsewhere [5], here we will focus on the formation of the antihydrogen beam.

## PULSED COLD ANTIHYDROGEN FORMATION

The antihydrogen beam formation will take places in AEGIS by a two step process: the first one is the production of very cold antihydrogen in Rydberg states and the second one is the acceleration in one direction (the horizontal one) of the Rydberg antihydrogen using inhomogeneous electric fields. A key point is that the antihydrogen formation will happen within a short time (of the order of 1 μs). This "pulsed production scheme" strongly differs from the existing schemes where the antihydrogen is produced during long time intervals (seconds in the ATHENA experiment) [8].

Cold antihydrogen in Rydberg states will be produced by the charge exchange reaction [5] [9]

$$Ps^* + \bar{p} \rightarrow \bar{H}^* + e^- \qquad (1)$$

between Rydberg positronium Ps* and cold (100 mK or below) antiprotons. Rydberg positronium (with principal quantum number n of the order of 20-30) will be formed by laser exciting the ground state positronium emitted by a porous material hit by a positron bunch ($10^8$ positrons with a time length of 10-20 ns).

The production of antihydrogen happens when the Rydberg positronium traverses the cold antiproton cloud. Taking into account that the velocity of the Rydberg positronium must be of the order of $10^4$-$10^5$ m/s to optimize the cross section for the reaction of Eq. 1 and that the antiproton cloud dimensions are of the order of a few mm, therefore the production time is defined within about 1 μs.

This pulsed production scheme of antihydrogen offers the possibility to easily measure the resulting antihydrogen temperature (by time of flight); it also provides a t=0 time for gravity measurement and it allows to form the antihydrogen beam as described below.

The trapped antiproton temperature is the main factor determining the resulting antihydrogen temperature provided that it is greater than about 100 mK: below this value kinematics effects in the charge exchange reaction are not completely negligible. FIGURE 1 (not to scale) shows the scheme of the region of the experiment where the antihydrogen will be formed, accelerated and sent through the grating system. Not shown in the drawing are the regions of the apparatus devoted to catch and cool antiprotons from the CERN Antiproton Decelerator and to accumulate positrons.

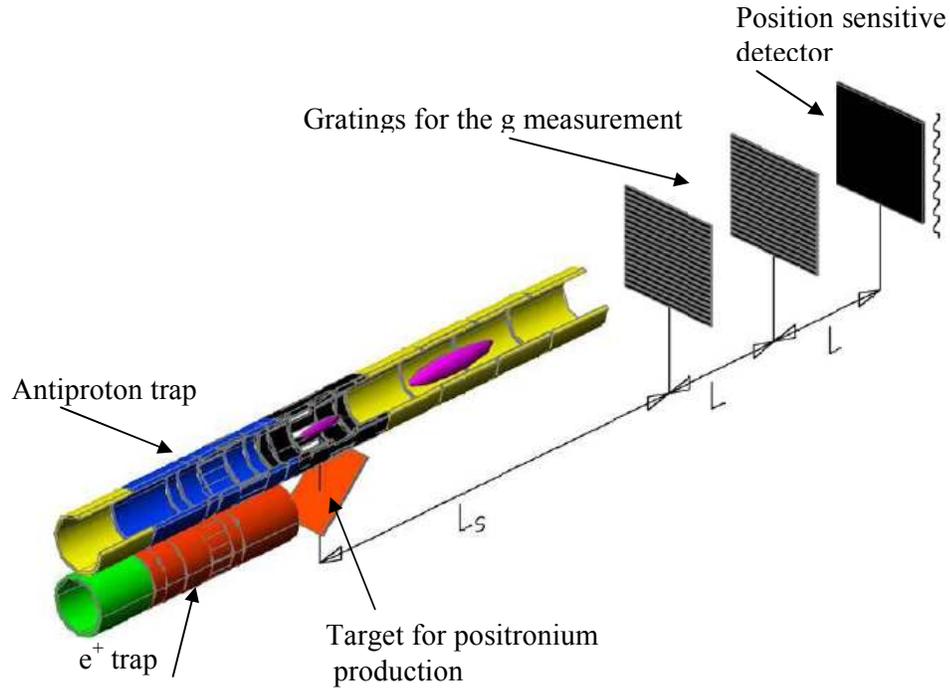

**FIGURE 1.** The figure (not to scale) shows the two parallel Penning-Malmberg traps that will be used in AEGIS to manipulate the antiprotons and positrons and to form and accelerate antihydrogen. They will be mounted inside a 100 mK cryostat in 1 Tesla magnetic field. The upper trap is devoted to antiprotons. Cold antiprotons wait for positronium in the black region. The lower trap is devoted to positrons: they will be sent to to the porous target mounted in front of the antiproton trap to produce positronium. Laser pulses will excite the positronium to selected Rydberg states having n in the range 20-30. The yellow region of the upper trap shows the bunch of antihydrogen after the acceleration (as described in the text). Finally the two material gratings followed by a position sensitive detector for the gravity measurement are shown. $L_s$ is 30 cm, L=40 cm. The trap radius is 8 mm. The gratings and detector dimensions are 20x20 cm$^2$.

## THE PENNING TRAP RYDBERG ACCELARATOR

The use of inhomogeneous electric fields to accelerate and manipulate Rydberg atoms has been proposed long time ago [10] but only recently this idea has become an experimental reality. In particular experimental demonstration of the acceleration, deceleration and even trapping has been achieved by members of AEGIS with hydrogen atoms in Rydberg states with quantum numbers very similar to those to be used in AEGIS [11].

To first approximation the energy levels E of the (anti)hydrogen atom in an external, homogeneous electric field of magnitude F are given, in atomic units, by

$$E = -\frac{1}{2n^2} + \frac{3}{2}nkF \qquad (2)$$

where k is a quantum number which runs from −(n − 1 − |m|) to (n−1−|m|) in steps of two and m is the azimuthal quantum number. Here 1 a.u. = 27.211 eV in the case of the energy and $5.14 \times 10^9$ Vcm$^{-1}$ for the electric field strength.

If the excited atoms are moving in a region where the amplitude of the electric field is changing then their internal energy changes according to Eq. 2 and thus, to conserve the total energy, they are accelerated or decelerated; the change of kinetic energy is 3/2 nk ΔF. This is described as a force acting on the atom by taking the space derivative of Eq. 1.

The experiments with hydrogen have shown that using time varying electric fields changes of velocity of the several hundreds m/s can be achieved in a few mm space using electric fields only limited by field ionization.

The formation of the beam of antihydrogen in AEGIS will be obtained by switching the voltages applied to the antiproton trap electrodes (immediately after the antihydrogen formation) from the usual Penning trap configuration to a new configuration that we call the "Rydberg accelerator".

To achieve this it is required to generate an electric field having an amplitude decreasing (or increasing) with z. Antihydrogen with positive (negative) k will be accelerated toward the system of gratings. The accelerating electric field will stay on for a selected time interval.

The gravity measurement procedure sets the requirements about the beam. The gratings of the Moiré device select specific trajectories of the atoms that can arrive on the detector. The distribution of the number of atoms arriving on the detector as a function of the vertical coordinate shows a periodical pattern due to the gratings. The gravity force causes a vertical shift of this pattern which depends on the time of flight T between the two gratings. Then to get g it is necessary to reconstruct, in addition to the vertical position on the detector, the horizontal velocity of the particles while they travel through the gratings. For this reason, it is important to have a start time well defined and a limited axial spread of the beam at the time when the accelerating electric field is switched off.

The transverse velocity of the beam does not influence the measurement result but it affects of course the number of particles arriving on the detector and the size of the gratings and detector itself. It is then mandatory to implement an acceleration procedure that does not heat the beam in a significant way in the radial direction.

We have simulated the expected velocity distribution in AEGIS taking into account the electric field obtained with the true geometry of the trap electrodes and realistic hypothesis about the initial antihydrogen spatial extension ( few mm radius and about 1 cm axial length ) and the expected state population.

The antihydrogen state population has been calculated by a dedicated CTMC code simulating the charge-exchange process in the AEGIS regime [5]. The Rydberg antihydrogen atoms will be produced not in a single quantum state but they will populate a distribution of states. The mean value of the principal quantum number $n_H$ of antihydrogen is related by the positronium one by $n_H \approx \sqrt{2} n_{Ps^*}$ but a distribution having a rms deviation of few units is expected.

As a reference the results here reported refer to a Gaussian distribution of $n_H$ having mean value 30 and a rms deviation equal to 4. For every n all the possible k have been included. In these conditions the product nk ranges from zero to about 1000.

The main consequence is that the various atoms will acquire different velocities in the same electric field and a beam with a large velocity spread will be obtained. To partially reduce the velocity spread it is important to use a time-varying electric field as shown in FIGURE 2.

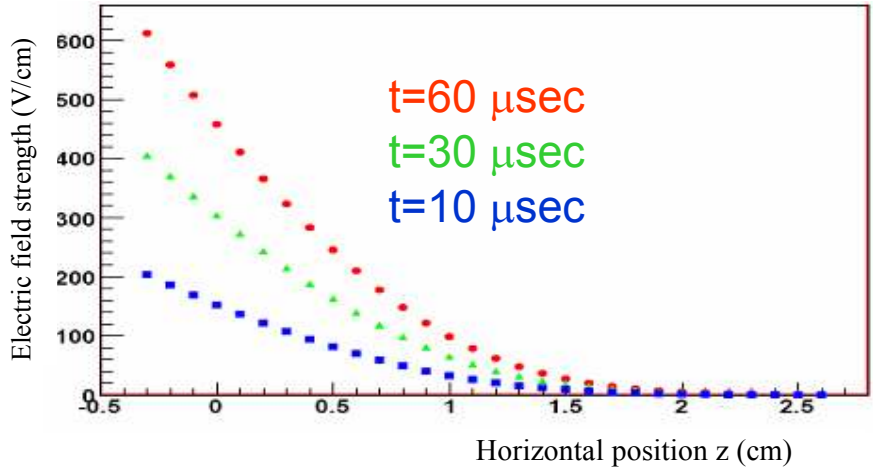

**FIGURE 2.** The figure shows the amplitude of the accelerating electric field on the trap axis versus z after 10, 30, 60 μs from the beginning of the acceleration process. The electric field stays on for 80 μsec in these simulations. It is obtained by applying voltages to radial sectors of the cylindrical trap electrodes. The accelerating electrodes are constituted by those of the antiproton trap shown in figure 1 together with some neighboring electrodes. These electrodes are split into four sectors (in the radial plane) with 135 degrees and 45 degrees angular extension. A voltage $V_0$ is applied to one of the big sectors while a voltage $-V_0$ is applied to the one opposite to it. The smaller sectors are grounded. The required distribution of electric field strength along the propagation direction of the beam is generated by applying a time-dependent potential to the relevant electrodes. The resulting electric field direction is almost perpendicular to the magnetic field. The antihydrogen cloud after production is centered at z=0.

FIGURE 3 shows the axial velocity before and after the acceleration for a cloud of antihydrogen produced with antiprotons cooled to 100 mK and distributed uniformly within 3 mm radius (that is within a large fraction of the 8 mm trap radius) and 8 mm length. Only 10% of the particles are lost on the trap electrodes. The corresponding distribution of the axial positions immediately after the accelerating electric field is switched off is shown in Fig 4: its spread is consistent with the requirements about the gravity measurement.

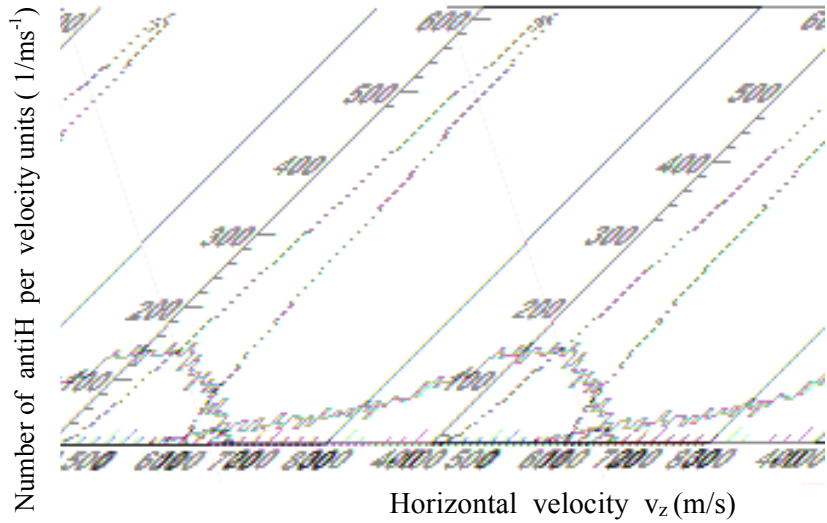

**FIGURE 3.** The red continuous curve is the distribution $dN/dv_z$ of the horizontal velocity at the time when the accelerating electric field is switched off. For comparison the dotted black is the initial velocity distribution (Maxwell-Boltzmann distribution at a temperature of 100 mK). The mean value of the red curve can be tuned by playing with the field intensity.

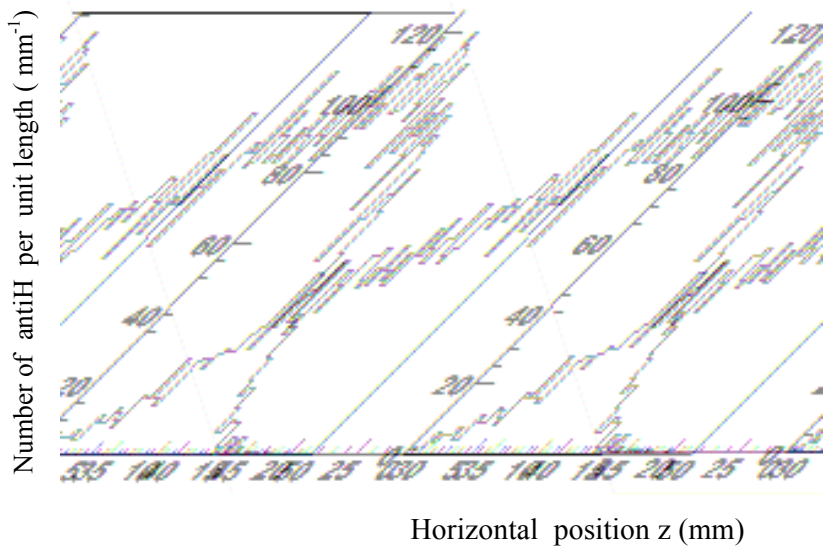

**FIGURE 4.** Distribution of the horizontal position of antihydrogen after acceleration. The plot refers to the same conditions as figure 3.

# TRANSVERSE VELOCITY DISTRIBUTION OF THE ANTIHYDROGEN BEAM

The amplitude of the radial electric field increases with the distance from the trap center thus providing an average deceleration force resulting in a reduction of the mean radial velocity. This is particular relevant for antihydrogen produced within 1-2 mm from the trap center. In the conditions of figure 2 (100 mK antihydrogen distributed within 3 mm radius) the distributions of the transverse velocity before and after the acceleration are practically identical. This is already a positive result because it indicates that, on average, the acceleration procedure is not destroying the efforts placed in producing ultracold antihydrogen. Particularly interesting if the comparison between the transverse velocity before and after the acceleration for antihydrogen initially produced within a smaller radius: 1.5 mm. In this case the net effect is radial cooling as shown in FIGURE 5.

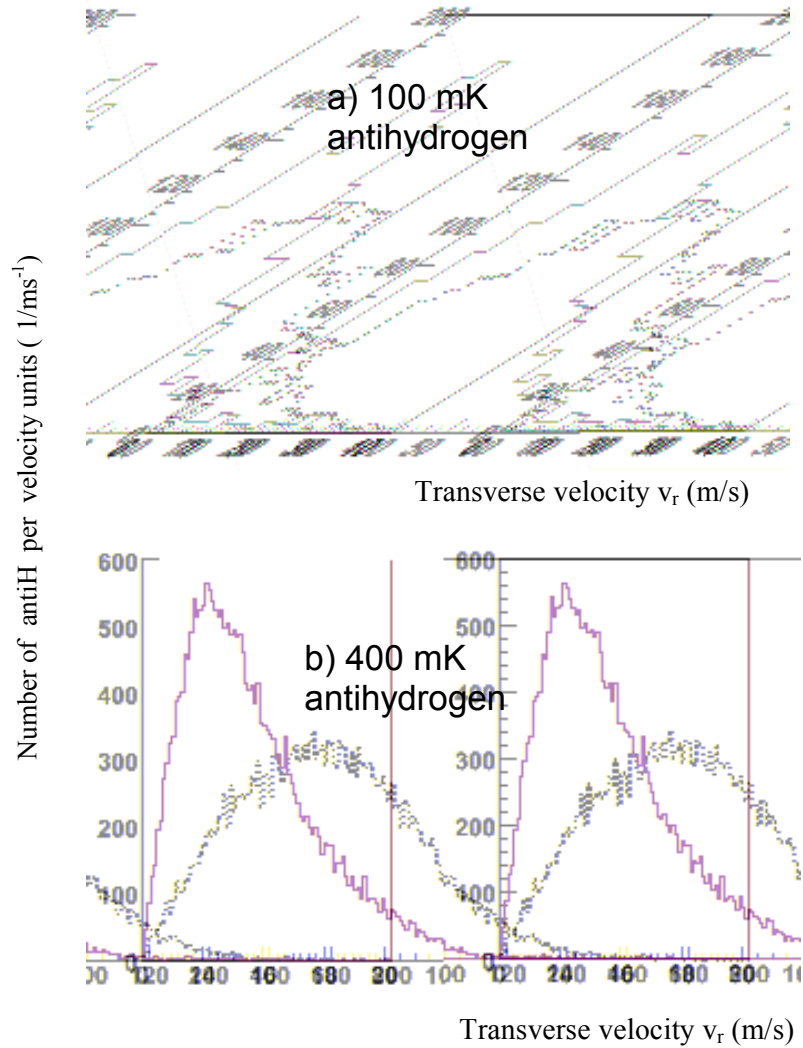

**FIGURE 5.** These two figures compare the transverse velocity distribution dN/dv$_r$ as a function of v$_r$ of antihydrogen before (black dotted plot) and after (red continuous plot) the acceleration. It is assumed that the antihydrogen is initially produced within a cylinder having 1.5 mm radius and 8 mm length. The initial velocity, before the acceleration, corresponds to a Maxwell distribution with 100 mK in figure a) and 400 mK in figure b). The cooling effect is evident in both cases.

# EFFECT OF THE MAGNETIC FIELD

The Stark acceleration of antihydrogen in AEGIS will happen in presence of the uniform magnetic field of the Penning-Malmberg traps. Under completely general conditions, the dynamics of Rydberg hydrogen atoms in electric and magnetic fields with arbitrary mutual orientation are rich and complex [12]. Several regimes are possible depending on the range of parameters (values of the electric and magnetic fields and principal quantum number values). The magnetic field in the antihydrogen formation region is chosen following a compromise between the requirement of the charged particle trap (that demands high magnetic fields) and the need to softly perturb the Rydberg antihydrogen. The choice of B=1 Tesla realizes this compromise for Rydberg antihydrogen having n not higher than about n=35. In this regime n is still a good quantum number and the diamagnetic correction is small (but not completely negligible) compared to the Zeeman one in the Hamiltonian and the classical dynamics is regular [13]. For larger n and higher magnetic field the classical dynamics could become chaotic and the quantum levels can show large avoided crossing [11]. These effects could influence the efficiency of the acceleration procedure. Detailed quantum calculations to exactly quantify these effects are in progress. The presence of the magnetic field is here accounted modifying Eq. 2 with the one giving the energy levels to the first order in the perturbation theory for perpendicular electric and magnetic fields [14]

$$E = -\frac{1}{2n^2} + n'\sqrt{\gamma^2 + 9n^2 F^2} \quad . \tag{3}$$

Here γ is the magnetic field in atomic units and n' is a quantum number assuming the values $-\left(\frac{n-1}{2}\right), -\left(\frac{n-3}{2}\right), \ldots \left(\frac{n-1}{2}\right)$ .

Figure 6 compares the simulated horizontal velocity distribution of figure 3 with the one obtained using the same electric field and the same initial conditions of the antihydrogen cloud but including the magnetic field. The effect is a reduction of the final velocity values which is tolerable. The radial cooling effect discussed in the previous paragraph is maintained in presence of magnetic field.

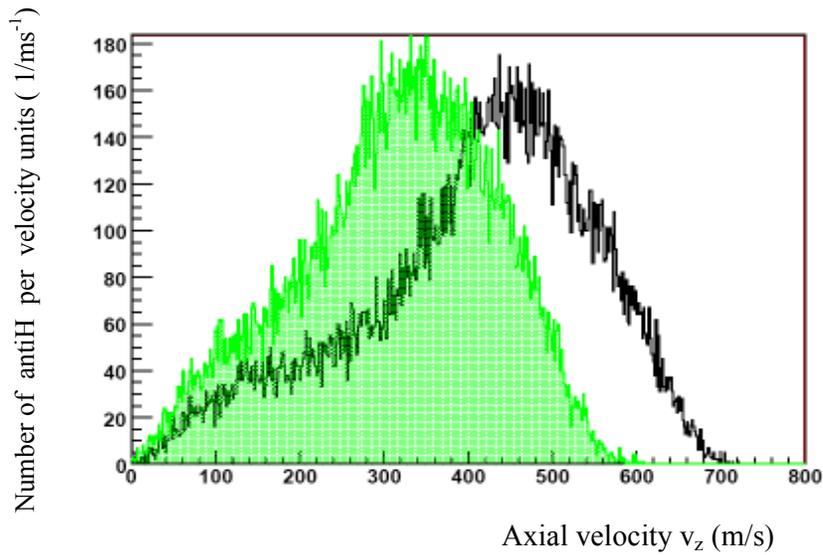

**FIGURE 6.** Simulated distributions of the horizontal antihydrogen velocity with (filled green plot) 1 Tesla magnetic field and without (black curve). The black curve is the one of figure 3. The initial antihydrogen cloud are the ones used to obtain figure 3 (a cloud of 100 mK temperature with 3 mm radial size and 8 mm length and having a mean value of n equal 30 with rms=4). For every n all possible values of n' with uniform distribution are considered.

## CONCLUSIONS

The recent developments about the manipulation of Rydberg hydrogen by using inhomogeneous electric fields together with the experience gained at the CERN AD to form antihydrogen are merged together in the design of AEGIS to form a beam of cold antihydrogen. The beam will be used to perform the first gravity measurement on antimatter. In the long term physics program of the collaboration several improvements are planned: here we mention the use of negative ions to better cool the antiprotons [15] and of quasi-continuous Lyman-$\alpha$ laser for cooling the formed antihydrogen beam [5,16]. Finally the possibility of accelerating (and decelerating) and thus transporting antihydrogen atoms opens the way to investigate novel antihydrogen trapping strategies in which the production and the trapping regions are spatially separated [5].